\documentclass[natbib209]{emulateapj}

\usepackage{amsmath}
\usepackage{epsfig}
\usepackage{color}

\newcommand{\pd}{\partial}

\newcommand{\bdot}{\mbox{\boldmath $\cdot$}}

\newcommand{\curl}{\mbox{\boldmath $\nabla \times$}}

\newcommand{\BB}{{\bf B}}
\newcommand{\JJ}{{\bf J}}

\newcommand{\uvr}{\mbox{\boldmath $\hat{r}$}}

\shorttitle{Helicity Reversals in a Convective Dynamo}

\shortauthors{Miesch, Zhang \& Augustson}

\begin{document}

\title{Magnetic Helicity Reversals in a Cyclic Convective Dynamo}

\author{Mark S. Miesch}
\affil{High Altitude Observatory, National Center for Atmospheric Research, Boulder, CO, 80307-3000, USA: miesch@ucar.edu}
\email{miesch@ucar.edu}

\author{Mei Zhang} 
\affil{Key Laboratory of Solar Activity, National Astronomical Observatories, Chinese Academy of Sciences, Datun Road A20, Chaoyang District, Beijing 100012, China}

\author{Kyle C. Augustson}
\affil{CEA/DRF/IRFU Service d'Astrophysique, CEA-Saclay, Batiment 709, 91191 Gif-sur-Yvette Cedex, France}

\begin{abstract}
We investigate the role of magnetic helicity in promoting cyclic magnetic activity in a global, 3D, magnetohydrodynamic (MHD) simulation of a convective dynamo.  This simulation is characterized by coherent bands of toroidal field that exist within the convection zone, with opposite polarities in the northern and southern hemispheres.   Throughout most of the cycle, the magnetic helicity in these bands is negative in the northern hemisphere and positive in the southern hemisphere.  However, during the declining phase of each cycle, this hemispheric rule reverses.  We attribute this to a global restructuring of the magnetic topology that is induced by the interaction of the bands across the equator.   This band interaction appears to be ultimately responsible for, or at least associated with, the decay and subsequent reversal of both the toroidal bands and the polar fields.  We briefly discuss the implications of these results within the context of solar observations, which also show some potential evidence for toroidal band interactions and helicity reversals.
\end{abstract}

\section{Introduction}\label{sec:intro}

Magnetic helicity, $H_m$, is a known agent of self-organization in turbulent magnetohydrodynamic (MHD) flows.  As an invariant of the ideal MHD equations, it can promote large-scale dynamo action by linking large and small scales.  This may occur in a self-similar manner, as in the inverse cascade of magnetic helicity in homogeneous MHD turbulence, or non-locally in spectral space, as in the turbulent $\alpha$-effect of mean-field dynamo theory \citep{brand05}.

The emerging magnetic flux that forms solar active regions is known to be helical in nature, suggesting that magnetic helicity plays an important role in the solar dynamo.  Photospheric and coronal observations reveal a systematic hemispheric rule, such that magnetic loops in the northern hemisphere (NH) tend to have a negative helicity while loops in the southern hemisphere (SH) tend to have a positive helicity \citep{pevts14}.  The hemispheric helicity rule is rather weak, with large scatter, but it is exhibited by 60--75\% of large active regions.  Other proxies for $H_m$ exhibit stronger hemispheric rules, such as the chirality of quiescent magnetic filaments \citep{pevts14}.

Several recent studies suggest that this hemispheric rule may reverse in the declining phase of the solar cycle, when the subsurface toroidal flux that gives rise to active regions is presumably diminishing \citep{tiwar09,hao11}.    In particular, \citet{hao11} studied a sample of 64 active regions spanning the declining phase of Cycle 23 and the rising phase of Cycle 24 (2006-2010).   They found that the data in the rising phase were consistent with the hemispheric helicity rule but that the data for the declining phase showed the opposite trend.  Again, both trends were weak (correlation coefficients $\lesssim 0.2$) but significant.   Of the 30 active regions studied during the declining phase of Cycle 23, only 8 (27\%) obeyed the hemispheric rule.

\begin{figure*}
\centerline{\epsfig{file=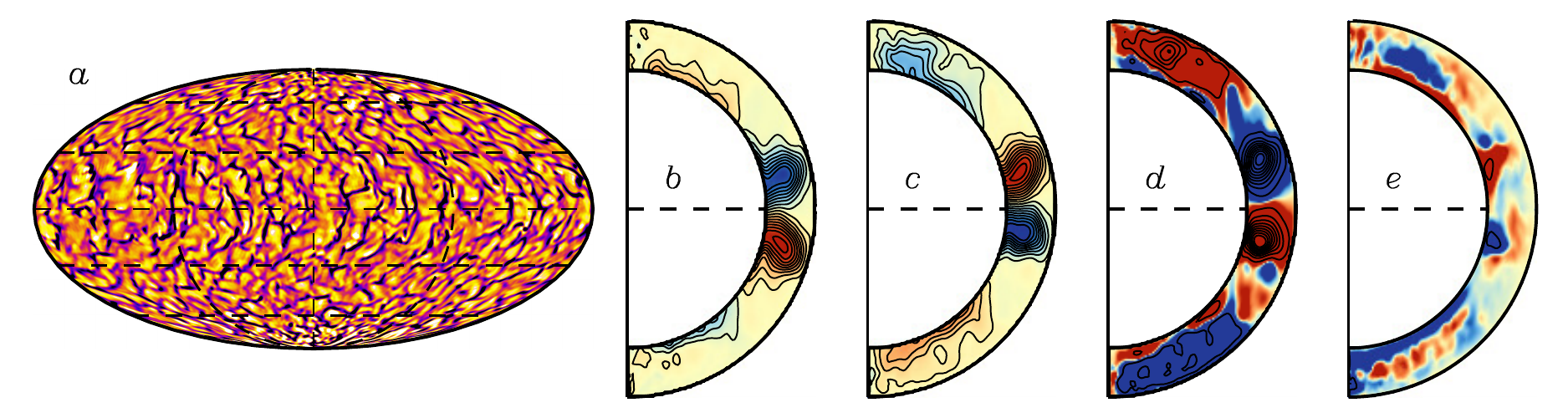,width=\linewidth}}
\caption{Illustrative results from Case K3S. ($a$) vertical velocity near the top of the convection zone ($r = 0.95$, $t=112$ yrs), shown in Molleweide projection (yellow upflow, blue downflow, saturation level $\pm 100$ m s$^{-1}$). Dashed lines indicate reference latitudes and longitudes. ($b$,$c$) Mean toroidal magnetic field $\left<B_\phi\right>$ at ($b$) cycle maximum ($t =$ 106.3 yrs) and ($c$) in the rising phase of the following cycle ($t =$ 109.2 yrs).   Red and blue denote eastward (prograde) and westward field respectively, with contours spanning $\pm $ 18 kG.  ($d$, $e$) Longitudinally-averaged magnetic helicity density $\left<h_m\right>$ (red positive, blue negative) at $t = $ 109.2 yrs for the ($d$) mean and ($e$) fluctuating field components.  The color table (identical in both) is heavily saturated to highlight sign changes but the peak magnitude in ($d$), 1.90 $\times 10^{17}$ G$^2$ cm, is about an order of magnitude larger than the peak magnitude in ($e$), 1.45 $\times 10^{16}$ G$^2$ cm.\label{fig:hfig}.}
\end{figure*}

In this paper we investigate the role of magnetic helicity in a global 3D MHD simulation of a convective dynamo.  This simulation exhibits regular magnetic cycles and is described in detail by \citet{augus15}, hereafter ABMT15.  We demonstrate that the polarity reversals that sustain the magnetic cycles in this simulation involve a restructuring of the magnetic topology that is reflected in and perhaps regulated by the evolution of the magnetic helicity.  A notable feature of this evolution is a weak helicity reversal in the declining phase of each magnetic cycle.

\section{Model and Analysis}

In this paper we consider a convective dynamo simulation K3S, which is described in detail in ABMT15.  The simulation is a result from the Anelastic Spherical Harmonic (ASH) code, which solves the three-dimensional (3D) MHD equations in a rotating spherical shell under the anelastic approximation \citep{clune99,brun04}.  In terms of its setup, this is a typical ASH simulation but with one exception; the explicit subgrid-scale (SGS) viscosity has been replaced by an implicit numerical diffusion that operates at the grid scale.  Meanwhile, the magnetic diffusion is similar to previous ASH simulations, with an explicit SGS magnetic diffusivity $\eta(r) = 8 \times 10^{12} (\rho_2/\rho)^{1/2}$, where $\rho_2$ is the density at the outer boundary.  This implicit large-eddy simulation (ILES) approach has enabled K3S to achieve a higher effective Reynolds number ($\sim$ 350) and a lower effective magnetic Prandtl number ($\sim$ 0.23) than in previous ASH simulations.  The magnetic Reynolds number is about 80.  The spatial resolution is $N_r, N_\theta, N_\phi = $ 200, 256, 512, corresponding to a maximum spherical harmonic degree of $\ell_{max} = 170$.

The computational domain extends from $r = 0.72$--$0.97 R$, where $R$ is the solar radius.  The entire domain is convective; there is no overshoot region and no tachocline.  The density contrast across the shell is 45 and the boundaries are impenetrable and stress-free, with a fixed heat flux (constant $\pd S/\pd r$, where $S$ is the specific entropy).  The inner boundary is assumed to be a perfect electrical conductor and the outer boundary is matched to an external potential field.

The rotation rate of the coordinate system $\Omega_0 = 7.8 \times 10^{-6}$ rad s$^{-1}$ is a factor of three larger than the solar rotation rate in order to promote and investigate the large-scale dynamo action induced by helicity and shear.  The Rossby number varies from about 0.12 to 0.33 depending on the phase of the magnetic cycle.  No mean flows are imposed.  Rather, they are established self-consistently by the convective momentum and energy transport.  The differential rotation is solar-like in the sense that the equator rotates about 30\% faster than the poles, but $\Omega$ contours are more cylindrical (aligned with the rotation axis) than the solar internal rotation inferred from helioseismology.

A striking feature of K3S is that it exhibits regular cycles with an average duration of $P_{cyc} \sim 3.1$ years, which corresponds to a full magnetic cycle period of 6.2 years.  Though this cycle period is significantly shorter than the 11-year solar cycle, the ratio of $P_{cyc}$ to the rotation period $P_{rot}$ is comparable (243 for Case K3S, 287 for the Sun).  This regular cycle has persisted throughout the entire simulation interval, which now exceeds 110 years.  However, it was interrupted temporarily by a {\em grand minimum} phase that lasted for about five cycles (15.3 years).  After the grand minimum, the regular cycle resumed.  For a detailed discussion of the cycle characteristics and the grand minimum see ABMT15.

Here we focus on the role of the magnetic helicity, which we define as
\begin{equation}\label{eq:Hm}
H_m = 2 \int_V h_m dV = 2 \int_V A B_r dV 
\end{equation}
where $h_m = A B_r$ is the magnetic helicity density and the integration proceeds over the computational volume $V$.  The toroidal and poloidal magnetic potentials $A$ and $C$ are defined by the Chandrasekhar-Kendall decomposition:
\begin{equation}\label{eq:CK}
\BB = \curl {\bf A} = \curl \left(A \uvr\right) + \curl \curl \left( C \uvr\right)  ~~~. 
\end{equation}
Throughout this paper we use spherical polar coordinates $r,\theta,\phi$ and $\uvr$ is the radial unit vector.  

\begin{figure*}
\centerline{\epsfig{file=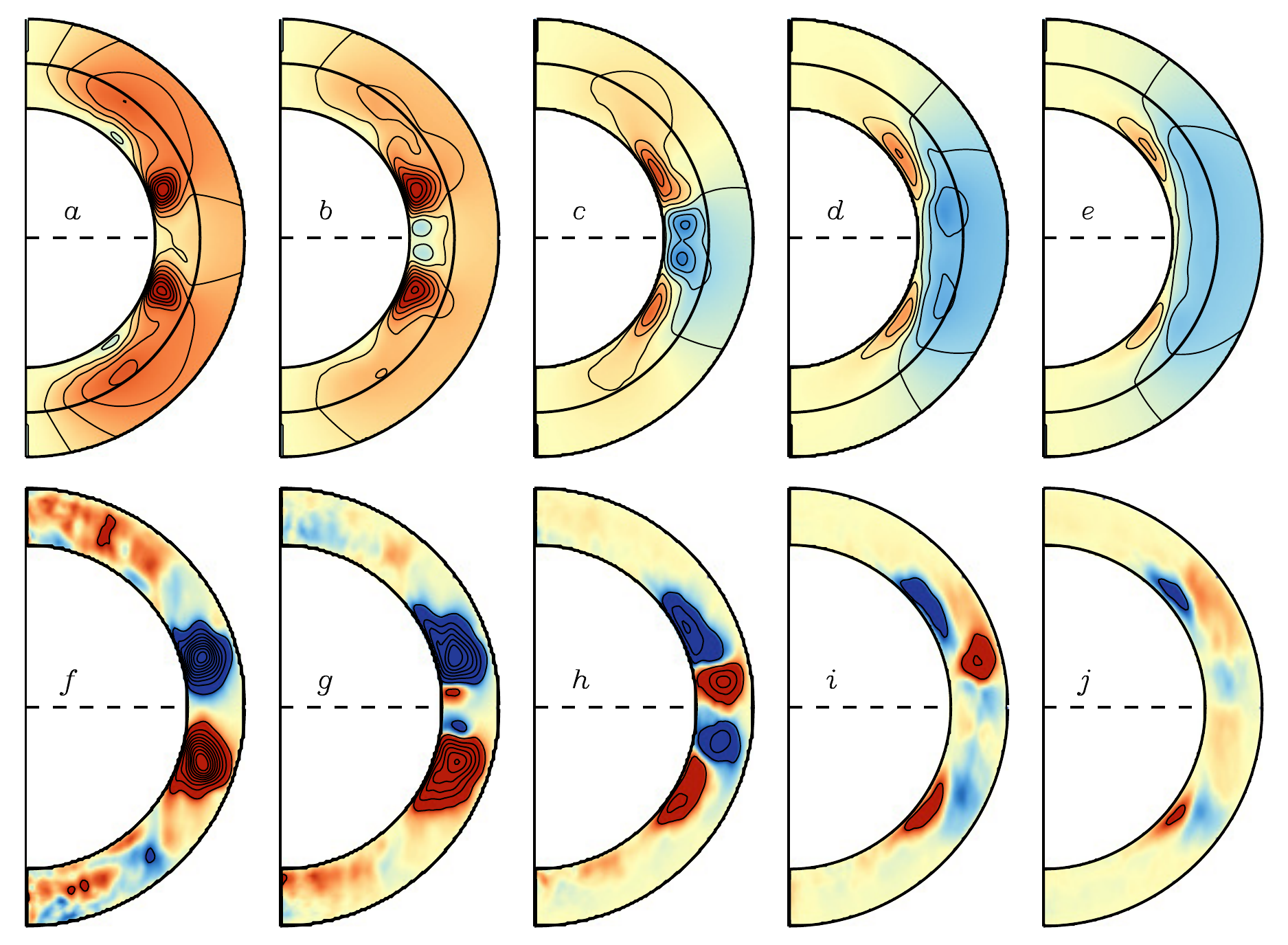,width=\linewidth}}
\caption{Mean poloidal field (top row) and magnetic helicity density $\left<h_m\right>$ (bottom row) at five times spanning the declining phase of a magnetic cycle: $t = $ 
($a$,$f$) 106.33 yr, ($b$,$g$) 106.54 yr, ($c$,$h$) 106.75 yr, ($d$,$i$) 106.96 yr, and ($e$,$j$) 107.17 yr.  In ($a$-$e$) red and blue denote clockwise and counter-clockwise field orientations respectively, with peak radial field strengths ranging from 6 kG in ($a$) to 0.6 kG in ($e$).  In ($f$--$j$) red and blue denote positive and negative helicity, with contours ranging between
$\pm 1.8 \times 10^{17}$ G$^2$ cm\label{fig:rev}.}
\end{figure*}

Even though there is a nonzero radial magnetic field at the outer surface of the computational domain, the helicity defined by eq.\ (\ref{eq:Hm}) is independent of the gauge chosen for the vector potential ${\bf A}$ \citep{berge85}.  Furthermore, for the special case of a spherical annulus, $H_m$ defined in this way is equal to the relative helicity, $H_R$, defined by \citet{berge84}.  For a proof that $H_m = H_R$ and a physical interpretation, see \citet{berge85}, \citet{low06}, and \citet{pevts14}.  In short, $h_m$ represents the toroidal component of ${\bf A}$, $A \uvr$, dotted into the poloidal component of $\BB$, $\curl \curl (C \uvr)$.   Since $A \uvr$ describes closed magnetic loops on a horizontal surface, the product $A B_r$ captures the full linkage of the poloidal and toroidal field components.

\begin{figure*}
\centerline{\epsfig{file=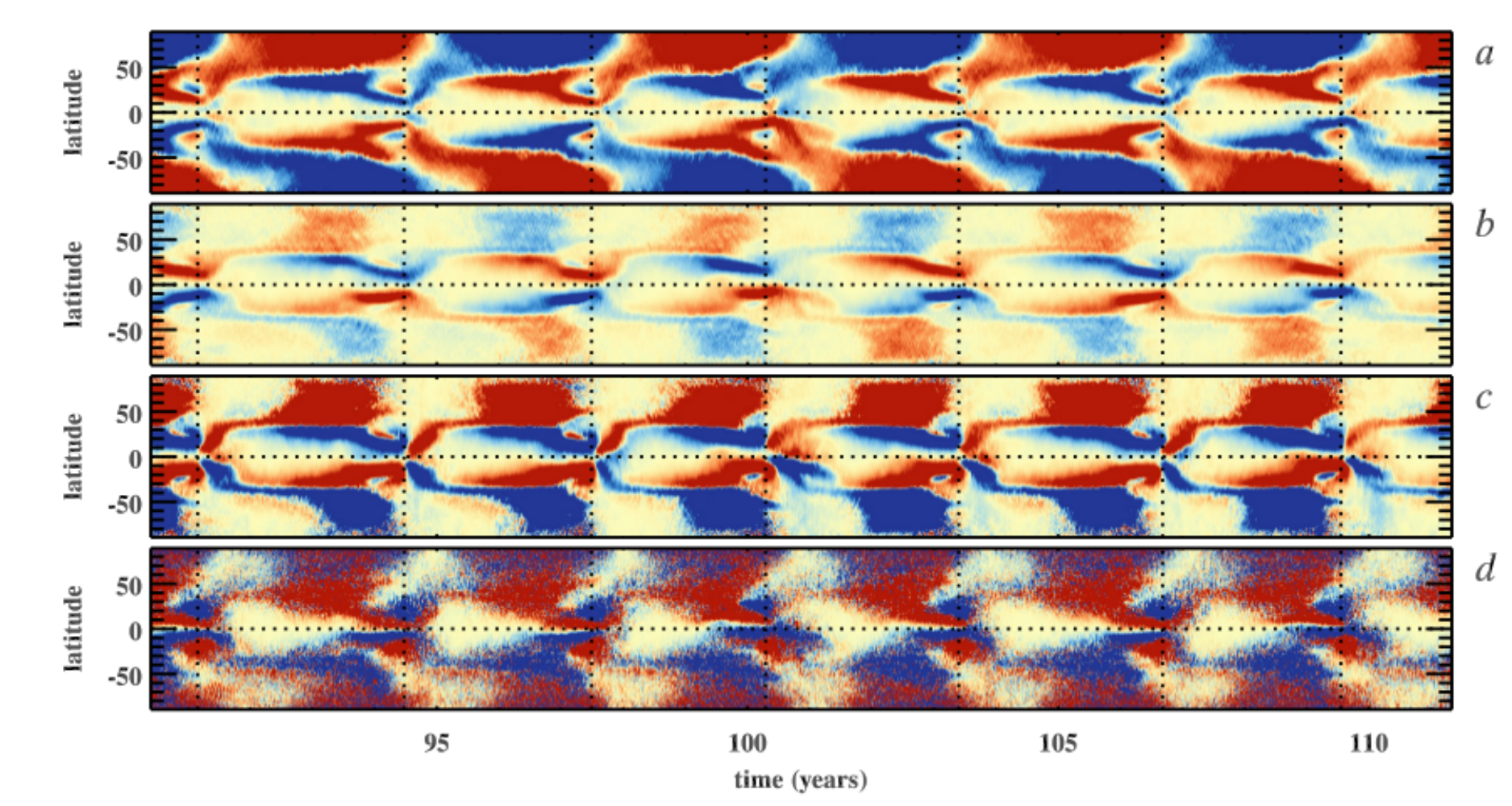,width=\linewidth}}
\caption{``Butterfly diagrams'' of zonally-averaged quantities versus latitude and time, spanning over six magnetic cycles:
($a$) $\left<B_r\right>$ at the outer boundary $r=0.97 R$, ($b$) $\left<B_\phi\right>$ at $r=0.94 R$, ($c$) $\left<h_m\right>$ at $r=0.94 R$, ($d$) $\left<h_c\right>$ at $r=0.94 R$.  Red and blue denote positive and negative values respectively and the color tables saturate at 
($a$) $\pm$ 200 G, ($b$) $\pm$ 2 kG, ($c$) $\pm 2 \times 10^{15}$ G$^2$ cm, and 
($d$) $\pm 10^{5}$ G$^2$ cm$^{-1}$. Vertical dotted lines indicate the times when the wreaths begin to decay, as shown in Fig.\ \ref{fig:bint}\label{fig:bfly}$a$.}
\end{figure*}

We will also be interested in the current helicity $H_c$ and its associated density $h_c$:
\begin{equation}
H_c = \frac{c}{4 \pi} ~ \int_V h_c dV = \frac{c}{4\pi} ~ \int_V \JJ \bdot \BB dV  ~~~.
\end{equation}
where $\JJ = \curl \BB$ is proportional to the electrical current density.  Unlike $H_m$, $H_c$ is not an invariant of the ideal MHD equations so it does not share the same topological properties (though these topological properties are compromised at the moderate magnetic Reynolds number of 80 employed here).  However, the vertical component of $h_m$, $J_r B_r$, is a common observational proxy used to quantify the helicity of solar magnetic fields \citep{pevts14}.

Figure \ref{fig:hfig} shows some highlights from Case KS3, including the structure of the convective motions ($a$) and the strong, coherent toroidal bands ($b$,$c$) that form within the convection zone (CZ) at low latitudes, despite the disruptive effects of turbulent mixing and magnetic buoyancy.  Following \citet{brown10} we will refer to these low-latitude bands as \textbf{magnetic wreaths} in order to distinguish them from the weaker, oppositely signed toroidal bands at high latitudes, as seen in Fig.\ \ref{fig:hfig}$c$.

Figs.\ ($b$) and ($c$) highlight the magnetic wreaths at cycle maximum, when the toroidal magnetic energy peaks, and during the rising phase of the following cycle, after the polarity of the wreaths has reversed.  Note the antisymmetry about the equator, which corresponds to a negative (dipolar) parity for both $\left<B_\phi\right>$ and $\left<B_r\right>$, as in the Sun.  Angular brackets denote averages over longitude.   Also shown in Fig.\ \ref{fig:hfig}$d$,$e$ are the mean and fluctuating components of the magnetic helicity density $\left<h_m\right>$ at the same instant as in Fig.\ \ref{fig:hfig}$c$.  We discuss these in the following section.

\section{Magnetic Helicity in A Cyclic Convective Dynamo}

In Fig.\ \ref{fig:hfig} we have decomposed the zonally-averaged magnetic helicity density into its ($d$) mean component $\left<A\right> \left<B_r\right>$ and its ($e$) fluctuating (non-axisymmetric) component $\left<(A-\left<A\right>)(B_r-\left<B_r\right>)\right>$.  During the rising phase of the cycle, as the wreaths are being built, these two components have opposite signs.  This demonstrates that the wreaths are helical in nature and suggests that the upscale spectral transfer of magnetic helicity from fluctuating to mean fields is playing an important role in their formation (at both low and high latitudes).  This interpretation is supported by considering the generation of magnetic helicity through ohmic dissipation (not shown), which is negative in the vicinity of the northern wreath and positive for the southern wreath during the rising phase of the cycle. However, the main source of both mean magnetic helicity and toroidal magnetic energy is the $\Omega$-effect.  

The net result from both contributions (spectral transfer and the $\Omega$-effect) is a magnetic helicity that is predominantly negative in the NH and positive in the SH, as inferred from solar observations.   This arises mainly from the mean field associated with the wreaths and is independent of the polarity of the cycle.  However, the mean helicity density at higher latitudes is reversed and shows a stronger anti-correlation with the fluctuating helicity (Fig.\ \ref{fig:hfig}$d$,$e$).

The magnetic helicity also appears to play an important role in the reversal of the toroidal and poloidal fields at the end of each cycle, as illustrated in Fig.\ \ref{fig:rev}.  Frames ($a$) and ($f$) show the mean poloidal field and magnetic helicity at cycle maximum, corresponding to the same time as the toroidal field in Fig.\ \ref{fig:hfig}$b$.  As time proceeds, a small ``bubble'' of oppositely-signed poloidal field appears at the equator near the base of the CZ (Fig.\ \ref{fig:rev}$b$).  This is associated with a sign reversal of $\left<h_m\right>$, producing a localized region of positive helicity in the NH and negative in the SH (Fig.\ \ref{fig:rev}$g$).  It is tempting to attribute the formation of this bubble to a magnetic helicity flux across the equator from south to north but a more comprehensive analysis (to be presented in a forthcoming paper) indicates that this is not the case.  Instead, cross-equatorial magnetic linkages between the wreathes promote a change in $\left<h_m\right>$ through resistive reconnection.  As time proceeds, this bubble rises through the CZ, amplifies, and unfurls toward higher latitudes, triggering a polar field reversal (Fig.\ \ref{fig:rev}$c$--$e$, $h$--$j$).

The nature of this time evolution suggests that the polar field reversal is a response to a change in the magnetic topology induced by the interaction of the wreaths across the equator.  This creates a magnetic bubble of opposite helicity that is set off from its environment by well-defined magnetic separatrices.  The subsequent evolution of this bubble occurs on an Alfvenic time scale, suggesting a magnetic relaxation process that occurs despite the high value of the plasma $\beta$ (10$^6$--10$^8$).  

\begin{figure*}
\centerline{\epsfig{file=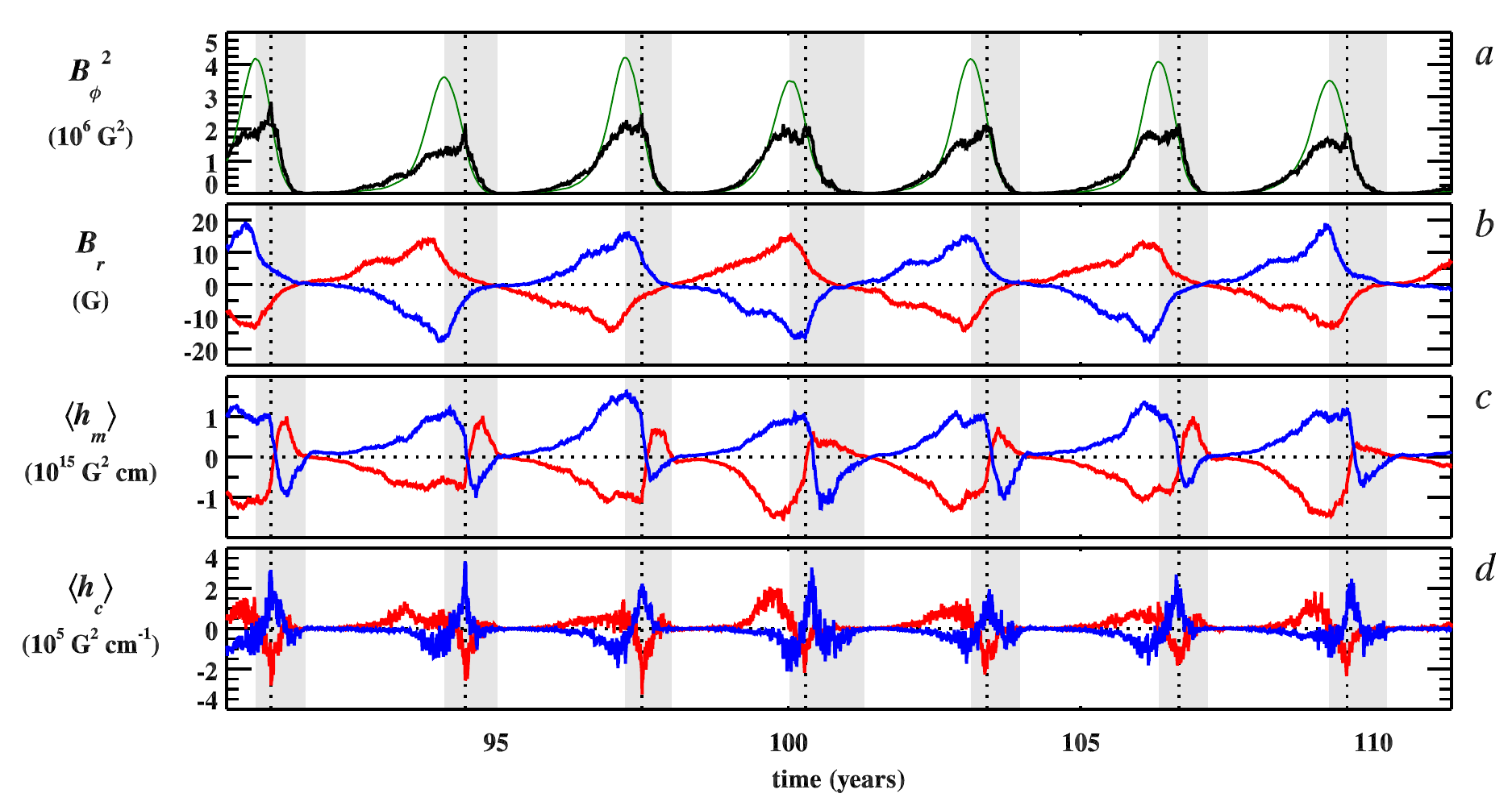,width=\linewidth}}
\caption{Time evolution of the integrated mean field and helicity. ($a$) Square of the mean toroidal field strength in the wreaths for $r = 0.94 R$ (black line) and integrated over all radii (thin green line), computed as described in the text.  The black line is multiplied by a factor of 10 to improve legibility.  The vertical dotted lines in all frames represent the times at which this black line begins to decay.  The grey areas represent the declining phase of the cycle, defined as the time between the maxima of the thin green line and the subsequent minima.  ($b$) Mean radial field $\left<B_r\right>$ at the outer surface, averaged over (red) the north pole (latitude $\geq 80^\circ$) and (blue) the south pole (latitude $\leq -80^\circ$). ($c$) Magnetic helicity density $\left<h_m\right>$ at $r = 0.94 R$, averaged over (red) 0$^\circ$--30$^\circ$ latitude and (blue) -30$^\circ$--0$^\circ$ latitude. ($d$) As in ($c$) but for the current helicity density $\left<h_c\right>$
\label{fig:bint}.}
\vspace{.15in}
\end{figure*}

To elaborate on this point we take a typical poloidal field strength at cycle max of $\sim$ 6 kG (Fig.\ \ref{fig:rev}$a$), and a density corresponding to the mid CZ, $\rho \sim 0.06$ g cm$^{-3}$. This give an Alfvenic time scale to cross the CZ of $\sim 0.3R/V_A \sim$ 0.05 yr, where $V_A = B (4\pi\rho)^{-1/2}$.  The subsequent unfurling of the poloidal field near the top of the CZ at the end of the cycle ($B \sim$ 0.6 kG, $\rho \sim 4\times 10^{-3}$ g cm$^{-3}$) could then proceed on a time scale of about $\pi R / (2 V_A) \sim$ 1.3 yr.  So, this suggests that the entire unfurling process, from the formation of the bubble to the polar field reversal could occur in roughly $1.35$ yr if it were due to magnetic restructuring induced by the Lorentz force.  This implies that the bubble would expand to mid-latitudes in about half that time, or about 0.675 yr.  Compare this to the time interval of 0.63 yr between frames $b$ and $e$ of Fig.\ \ref{fig:rev}.

Fig.\ \ref{fig:bfly} shows the time evolution in more detail.  The butterfly diagram for $\left<B_\phi\right>$ in the upper CZ (Fig.\ \ref{fig:bfly}$b$) shows the slight equatorward propagation emphasized by ABMT15.  When the wreaths meet at the equator, it triggers a helicity reversal (Fig.\ \ref{fig:bfly}$c$) as seen in Fig.\ \ref{fig:rev}$g$.  This new helicity propagates poleward in conjunction with a new high-latitude toroidal band (Fig.\ \ref{fig:bfly}$b$,$c$).  Within 1-2 yr of the helicity reversal at $r = 0.94R$, the polar fields at the surface reverse (Fig.\ \ref{fig:bfly}$a$).  This is due to the poleward migration of oppositely signed radial flux from lower latitudes.  There is also a weak, localized, transient loop of opposing radial field that threads through the surface at low latitudes just before the global poloidal and toroidal fields reverse (also seen in ABMT15, Fig.\ 3).

Also shown in Fig.\ \ref{fig:bfly}$d$ is the mean current helicity density $\left<h_c\right>$, at the same radial level as the magnetic helicity in Fig.\ \ref{fig:bfly}$c$.  We find the current helicity to be much less structured than the magnetic helicity, exhibiting mixed polarity in both hemispheres with a less pronounce hemispheric asymmetry.  Furthermore, we find that the current helicity is a poor proxy for the magnetic helicity, particularly within the wreaths where $\left<h_c\right>$ and $\left<h_m\right>$ often have the opposite sign (Fig.\ \ref{fig:bfly}$c$).

The vertical dotted lines in Fig.\ \ref{fig:bfly} mark the times when the wreaths in the upper CZ begin to decay.  To quantify this, we first integrate the mean toroidal field $\left<B_\phi\right>$, over a latitude range of $0$--$30^\circ$ and $-30$--$0^\circ$ and then either integrate over radius (thin green line in Fig.\ \ref{fig:bint}$a$) or sample the field at $r = 0.94$ (black line), the same level as shown in Fig.\ \ref{fig:bfly}$b$--$d$.  We then normalize these fluxes with the effective area to give a mean toroidal field strength for the northern and southern wreathes.  We then square these two components and add them.  The times indicated by the vertical dotted lines coincide with the appearance of the magnetic bubble in Fig.\ \ref{fig:rev}$b$,$g$, which is seen as a prominent, repeating, quadrupolar pattern in the butterfly diagram of Fig.\ \ref{fig:bfly}$c$.  This suggests that the magnetic restructuring triggered by the interaction of the wreaths is responsible for their subsequent demise.  Similar cross-equatorial interactions among helical toroidal field structures have been reported by \citet{mitra10} and \citet{brown11}.

We define the declining phase of each magnetic cycle as the time interval when the mean toroidal magnetic energy in the wreaths is decaying, as quantified by the thin green line in Fig.\ \ref{fig:bint}$a$.  Note that the phasing of the polar field reversals is not solar-like, in the sense that they occur near cycle minima.  As shown in Fig.\ \ref{fig:bint}$c$, the declining phase coincides with a weak reversal in the magnetic helicity.   

The current helicity also exhibits a weak reversal during the declining phase of each cycle (Fig.\ \ref{fig:bint}$d$).  This precedes the reversal of $\left<h_m\right>$ and helps to bring it about, since the time derivative of $h_m$ involves a term that is proportional to $-\eta h_c$ \citep{brand05,mitra10}.  Thus, the generation of negative $h_c$ in the NH promotes the generation of positive $h_m$ through ohmic dissipation (vice versa for the SH).  This also accounts for why the signs of $\left<h_c\right>$ and $\left<h_m\right>$ are often opposite.

\section{Conclusion}

We have demonstrated that the cyclic magnetic reversals in a convective dynamo simulation are linked to a global restructuring of the magnetic topology as reflected by the magnetic helicity.  This global restructuring is triggered by the cross-equatorial interaction of toroidal magnetic bands (wreaths) that give birth to an axisymmetric magnetic bubble that is topologically disconnected from it surroundings.  The subsequent expansion and poleward migration of this bubble appears to be responsible for, or at least associated with, the diminishing of the toroidal bands and the reversal of the poloidal fields.  

Throughout most of the cycle, the magnetic helicity is negative in the NH and positive in the SH.  However, in the declining phase of the cycle, this hemispheric rule briefly reverses, due to the formation of the magnetic bubble.  This is reminiscent of solar observations, which show a similar hemispheric rule and possible evidence for a helicity reversal in the declining phase of the cycle (Sec.\ \ref{sec:intro}).  Thus, if this observed helicity reversal is confirmed, our results suggest that it may arise from the interaction of subsurface toroidal bands.

However, estimates of the magnetic helicity from photospheric measurements are sometimes based on the vertical component of the current helicity density, $B_r J_r$.  In our simulation, we find that the current helicity is a poor proxy for the magnetic helicity; $\left<h_c\right>$ is much less structured than $\left<h_m\right>$, with mixed polarity in both hemispheres.  This may account for why the observed hemispheric helicity rules are weak, with large scatter.

Other possible evidence for interactions among toroidal bands in the solar convection zone or tachocline has been recently reported by several authors based on close scrutiny of the sunspot butterfly diagrams and related short-term solar variability \citep{mcint15,camer16,munoz16}.  Continued work on toroidal band interactions and the role of magnetic helicity from both an observational and a modeling perspective promises to provide further insights into the subsurface dynamics that give rise to the solar cycle.

\acknowledgements
We thank the anonymous referee and Kevin Dalmasse for constructive comments on the manuscript.  M.\ Zhang acknowledges the support of the National Natural Science Foundation of China (Grants No. 11125314 and No. U1531247) and the Strategic Priority Research Program (Grant No. XDB09000000) of the Chinese Academy of Sciences.  Computational resources were provided by NASA's High End Computing (HEC) Program and the NCAR/Wyoming Supercomputing Center (NWSC).  The National Center for Atmospheric Research is sponsored by the National Science Foundation.


\end{document}